\documentclass{egpubl}
\usepackage{eurovis2026}
\usepackage{xspace}
\usepackage{xcolor}
\usepackage{kotex}
\usepackage{multirow}
\usepackage{booktabs}
\usepackage{tabularray}
\newcommand{\system}{\textit{Symetra}\xspace}

\newcommand{\edit}[1]{#1}

\usepackage{tcolorbox}

\newcommand{\val}[1]{\textit{\textsf{\footnotesize #1}}}
\newcommand{\g}[1]{\textit{#1}}

\SpecialIssuePaper         

\CGFccby

\usepackage[T1]{fontenc}
\usepackage{dfadobe}  

\usepackage{cite}  
\BibtexOrBiblatex
\electronicVersion
\PrintedOrElectronic

\usepackage{egweblnk}

\title[\system: Visual Analytics for the Parameter Tuning Process of Symbolic Execution Engines]%
      {\system: Visual Analytics for the Parameter Tuning Process of Symbolic Execution Engines}

\author[D. Hong, M. Kim, S. Cha \& J. Jo]
{\parbox{\textwidth}{\centering D.\,Hong\orcid{0009-0005-0890-2556},
        M.\,Kim\orcid{0009-0004-7097-3171},
        S.\,Cha\orcid{0000-0002-4697-8536},
        and J.\,Jo\orcid{0000-0002-5207-6010}
        }
        \\
{\parbox{\textwidth}{\centering Sungkyunkwan University, Republic of Korea}
}
}

\begin{document}


\maketitle
\begin{abstract}
%
%
Symbolic execution engines such as KLEE automatically generate test cases to maximize branch coverage, but their numerous parameters make it difficult to understand the parameters' impact, leading the user to rely on suboptimal default configurations. While automated tuners have shown promising results, they provide limited insights into why certain configurations work well, motivating the need for Human-in-the-Loop approaches.
In this work, we present a visual analytics system, \system, designed to support Human-in-the-Loop parameter tuning of symbolic execution engines.
To handle a large number of parameters and their configurations, we provide two complementary overviews of their impact on branch coverage values and patterns.
Building on these overviews, our system enables collective analysis, allowing the user to contrast groups of configurations and identify differences that may affect branch coverage.
We also report on case studies and a Human-in-the-Loop tuning process, demonstrating that experts not only interpreted parameter impacts and identified complementary configurations, but also improved upon fully automated approaches in both branch coverage and tuning efficiency.

\ccsdesc[500]{Human-centered computing~Visual analytics}
\ccsdesc[300]{Software and its engineering~Software testing and debugging}
\printccsdesc   


\end{abstract}  

\section{Introduction}
One important research question in software engineering is how to ensure that a program is robust to arbitrary inputs. This is typically accomplished by generating test cases, executing them, and checking for uncaught exceptions or violated assertions.
Symbolic execution~\cite{dart, exe, cute, 10.5555/1855741.1855756} is a powerful method that automatically generates such test cases. 
The key idea is to systematically explore the program's execution paths by treating input variables as symbolic values and generating inputs that exercise symbolic branches. 
Various symbolic execution engines, such as KLEE~\cite{10.5555/1855741.1855756}, Crest~\cite{10.1109/ASE.2008.69}, or Sage~\cite{10.1145/2093548.2093564}, have been researched and adopted.

A critical challenge in effectively leveraging these engines lies in configuring the appropriate parameters that influence their behavior and thus branch coverage. 
Modern engines incorporate diverse, independent techniques, such as search strategies~\cite{9507083,10.1145/3460120.3484813} and constraint solving~\cite{10.1145/3324884.3416589, partial} that come with numerous tunable parameters. As a result, it has ironically become more difficult to use the engines effectively due to the large number of parameters.
For instance, KLEE~\cite{10.5555/1855741.1855756}, a widely used symbolic execution engine, offers tens of tunable parameters spanning binary, continuous, and nominal types.
This creates a complex, high-dimensional search space with non-linear interactions between parameters, making the tuning process challenging. 
Furthermore, a single configuration may not be optimal in terms of branch coverage; the user often requires a set of complementary configurations (i.e., configurations whose test cases cover different sets of branches) to further improve branch coverage. For example, a configuration with the Breadth-First Search (BFS) strategy may provide high overall branch coverage, while one with a Depth-First Search (DFS) strategy may better explore deeply nested branches. 

In this work, we aim to enable the user to effectively leverage symbolic execution engines by providing a visual analytics approach to the parameter tuning process.
We observed that most users rely on command-line interfaces and general spreadsheet software to manually sift through thousands of configurations.
Although automated tuners for the engines~\cite{10.1145/3460120.3484813, 9507083, 10.1145/3510003.3510185} have shown promising results, our domain analysis revealed four critical limitations of these fully automated approaches.
First, the user rarely develops an understanding of why certain configurations work well, which would help them use the engines more efficiently in the future, such as when testing newer versions of the same software.
Second, even though a small number of configurations can achieve high branch coverage if they target different parts of the program (i.e., complementary configurations), the discovery of such configurations occurs only by chance and is not well understood.
Third, although existing tuners are designed to balance exploration and exploitation, the complex search space often leads to early saturation in branch coverage, i.e., repeatedly generating configurations that test similar code regions. Human guidance can help steer the search toward unexplored regions and break such saturation.
Finally, certain configurations can result in testing failures, yielding zero branch coverage and wasting computational resources (e.g., running for several minutes without testing any branch). However, existing automated approaches do not provide meaningful insights into their causes.
These limitations motivate Human-in-the-Loop approaches where visual analytics bridges automated optimization and human understanding.

To tackle these challenges, we present \system, a visual analytics system that helps the user analyze and understand the parameter tuning process for symbolic execution engines.
We demonstrate that the tuning process can be enhanced not only in a qualitative sense (i.e., providing insights that guide further improvements) but also in a quantitative sense (i.e., achieving higher branch coverage faster).
To our knowledge, \system is the first visual analytics system designed specifically for the tuning process of symbolic execution engines, addressing a critical need in the software testing community where visual analytics has been largely absent.
In collaboration with two domain experts, we identified user tasks and requirements that guided our design study.
We validated our design through three complementary evaluations: qualitative case studies and expert interviews to assess the system's analytical capabilities, and a quantitative evaluation demonstrating that Human-in-the-Loop parameter tuning improves fully automated approaches.

The contributions of this paper are:
\begin{enumerate}[nosep]
    \item Design and development of \system, the first visual analytics system for Human-in-the-Loop parameter tuning of symbolic execution engines,
    \item Abstraction and characterization of the domain situation in symbolic execution-based software testing, and 
    \item Qualitative and quantitative evaluation of \system through case studies, expert interviews, and a quantitative demonstration of Human-in-the-Loop parameter tuning.
\end{enumerate}

\begin{figure*}[ht]
  \centering
  \includegraphics[width=0.86\linewidth]{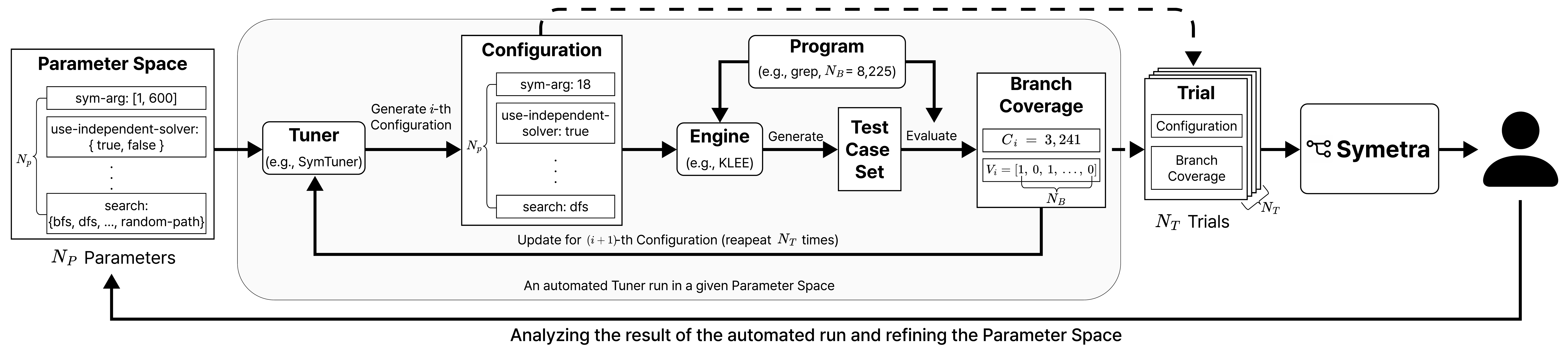}
  \caption{ \edit{\textbf{Human-in-the-Loop Parameter Tuning.}}
The gray box represents an experiment where a tuner iteratively optimizes configurations to maximize the branch coverage of the given program. The results of an experiment are trials, pairs of a configuration and its branch coverage, represented as branch coverage vectors ($V_i$) and branch coverage values ($C_i$), which are visualized in \system.
\edit{The user then analyzes the results and refines the parameter space for subsequent experiments.}
  }
  \label{fig:workflow}
  \vspace{-5mm}
\end{figure*}

\vspace{-2mm}

\section{Related Work}

In this section, we survey relevant research and systems from both information visualization and software engineering.

\vspace{-3mm}

\subsection{Parameter Tuning for Symbolic Execution}

Parameter tuning in symbolic execution engines has been a critical challenge since they became a mainstream testing technique~\cite{10.1145/800027.808445, 10.1145/3182657, 10.1145/360248.360252}. 
While symbolic execution engines such as KLEE~\cite{10.5555/1855741.1855756} have become widely adopted in academia and industry~\cite{10.1145/1985793.1985995, 10.1145/2408776.2408795}, maximizing their performance remains challenging due to the complex parameter space that must be tuned for each target program~\cite{10.1145/2509136.2509553, 10.1145/3180155.3180251, 7194611}.

Early parameter tuning approaches were mostly manual and relied heavily on default settings. For instance, many researchers simply adopted KLEE's default parameter values that were originally tailored for GNU Coreutils~\cite{10.1145/3368089.3409755, chen_et_al:LIPIcs:2018:9211, 10.1145/3238147.3238179, 10.1145/3293882.3330554, 7194611 }, even when testing entirely different programs. 
Others attempted to manually tune specific parameters while keeping the rest at their default values~\cite{10.1145/3324884.3416589, 10.1145/2509136.2509553, 10.1145/3092703.3092728, 10.1145/3180155.3180251, 7835264}, but this approach proved suboptimal given the enormous search space created by the combination of parameters, many of which are non-boolean.

Recent research has made progress toward automated parameter tuning. SymTuner~\cite{10.1145/3510003.3510185} represents a significant advance by introducing an adaptive learning algorithm that automatically adjusts parameter values during symbolic execution. The system evaluates performance with sampled parameter values and refines probability distributions of the sample spaces based on the results. 
Similar approaches, such as ParaDySE~\cite{9507083} and Learch~\cite{10.1145/3460120.3484813}, focus on tuning a specific parameter of the engines (i.e., search strategy) while leaving tens of other important parameters for the user to tune manually.
While these automated approaches are effective in improving branch coverage~\cite{10.5555/1855741.1855756, 8952548}, the lack of visual support makes it difficult for the user to understand the tuning process and derive insights for future improvements. To address this gap, we present \system, a system that complements fully automated approaches by making the tuning process more transparent.

\vspace{-5mm}

\subsection{Visual Analytics Systems for Parameter Tuning}
Visual analytics tools for (hyper)parameter tuning have been extensively studied, particularly in the context of hyperparameter optimization (HPO) for machine learning models. 
Many of these systems focus on supporting visual analytics to facilitate understanding of parameter configurations tested during the process.
For instance, HyperTendril~\cite{9222338} enables visual exploration of search results from multiple HPO algorithms. 
A wide range of visualization techniques has been introduced in this context, such as parallel coordinates between parameters representing a configuration as a polyline~\cite{10.1145/3097983.3098043, nni2021, 9222338, 10.1145/3292500.3330701, DBLP:journals/corr/abs-1807-05118}, scatterplots with dimensionality reduction~\cite{https://doi.org/10.1111/cgf.14300} to provide an overview of configurations across performance metrics and simpler, more familiar visualizations such as bar charts~\cite{10.1145/3290605.3300911, 10075647}.
Recent advancements have shifted towards interactive tuning, emphasizing a tighter integration between visual interfaces and search algorithms for human steerability. For example, ATMSeer~\cite{10.1145/3290605.3300911} allows the user to interactively refine the search space during the tuning process, providing greater control.

However, these systems cannot be directly applied to the parameter tuning process in software engineering.
The first reason is scalability. \edit{Prior} systems are typically designed to handle a relatively small number of parameters and configurations, often fewer than ten parameters and hundreds of configurations. This choice may stem from the long time required to evaluate each configuration. 
In contrast, modern symbolic execution engines have tens of parameters, and each configuration can be tested within a few minutes, leading to thousands of configurations per experiment. Consequently, scalability becomes critical; for instance, using parallel coordinates, as employed in prior work, may not be effective due to the high dimensionality and number of configurations.

The second reason is the unique domain situation of software testing, which we elaborate on in Section \ref{sect:domain}. Unlike in machine learning, where configurations produce a single performance metric, e.g., accuracy, the configurations in software testing produce a vector with each dimension indicating whether a specific branch was tested, requiring a tailored visualization interface.
Furthermore, it is possible to merge test cases generated by different configurations to complement one another to improve branch coverage, an important user task that is infeasible in machine learning and, therefore, not supported by previous systems.

In parallel, a few studies in software engineering~\cite{9161524, 9925664, 7102631} have proposed visualization systems to aid in understanding and navigating symbolic execution paths and states.
For instance, SymNav~\cite{9161524} introduces a visual analytics approach to assist security analysts in exploring symbolic execution traces, using parallel coordinates and control flow graphs to visualize critical paths in the execution tree.
Although these studies offer insights into understanding execution traces and paths, they do not address the challenges associated with parameter tuning for symbolic execution engines.
To our knowledge, \system is the first visual analytics system specifically designed for parameter tuning of symbolic execution engines, addressing a gap where neither general visualization tools nor existing HPO visualization systems can adequately support the unique requirements. 

\vspace{-4mm}

\section{Domain Analysis}\label{sect:domain}
We collaborated with two researchers from a university for ten months: E1 with 11 years and E2 with 3 years of experience in symbolic execution. 
\edit{We held monthly meetings where we observed their workflows, discussed their goals and tasks, and iteratively refined them into abstract forms. Each meeting, we presented updated prototypes, collected feedback on both the task abstraction and the visual design, and revised accordingly.}
\vspace{-4mm}

\subsection{Background and Terminology}

In software engineering, the quality of a test case set is assessed by counting the number of program branches executed by at least one test case in the set, a metric referred to as its \textbf{branch coverage}; note that a branch may be exercised multiple times by different test cases in the set, but it is counted only once.
As manually crafting a test case set requires a significant amount of time and labor, researchers have developed an automated technique called \textbf{symbolic execution}, which analyzes a program, treating the input variables as symbols, and generates the input that makes a certain part of the program executed~\cite{10.1145/360248.360252}.
KLEE~\cite{10.5555/1855741.1855756} is one of the most widely used symbolic execution engines that has been actively researched in recent years (hereafter, \textbf{engine}).
From a high-level view, it takes two inputs: \textbf{program} and \textbf{parameters}.
A program is specified as source code, and parameters are configurable options that determine the engine's behavior when generating test cases.
For example, a \emph{search} parameter determines how the engine traverses branches, which could be one of \val{bfs} or \val{dfs}.

It is known that choosing the proper parameters plays an important role in improving the branch coverage of the generated test set, which may be different program-by-program~\cite{9507083, 10.1145/3510003.3510185}.
To automate this parameter tuning process, researchers, including our domain experts, have focused on developing a tuner program (hereafter, \textbf{tuner}) that searches for parameters that maximize branch coverage for a given program.

A tuner optimizes the parameters by interacting with the engine in an iterative manner, which is depicted in the gray box of Figure~\ref{fig:workflow}.
The tuner first randomly samples parameter values from the given parameter space. 
Then, it runs the engine with those parameter values for a fixed time, generating a test case set.
Finally, a branch coverage of the generated set is evaluated, and the tuner updates its internal state to better sample parameter values.
This process is repeated for a certain time or iterations.
In a sense, the tuning process can be understood analogously to HPO in machine learning; our domain experts use an automated tuner (an HPO algorithm) that maximizes the branch coverage (accuracy) of the generated test case set (trained model) for a specific program (dataset).

We define an \textbf{experiment} as a single run of a tuner for the given target program and parameter space with $N_P$ parameters.
Figure~\ref{tab:param} shows an example of eight parameters.
In our study, we used a state-of-the-art algorithm, SymTuner~\cite{10.1145/3510003.3510185}, as a tuner. 
An experiment runs the engine multiple times, and we call each run \textbf{a trial}.
We denote a trial by an integer $i \in [1, N_T]$, where $N_T$ represents the total number of trials in the experiment.
For each trial, the tuner determines the values of the parameters, which we define as a \textbf{configuration}; a configuration is a key-value dictionary of size $N_P$ where keys are parameter names and values are the parameter values, e.g., $\{``sym\_arg":100,  ``search": ``bfs", ...\}$.
After the engine is run on a configuration, the tuner evaluates the branch coverage of the generated test case set.
The branch coverage is represented in two ways: as \textbf{a branch coverage vector} and \textbf{a branch coverage value}.
The branch coverage vector of trial $i$, denoted as $V_i\in\{0, 1\}^{N_B}$, is a binary vector of length $N_B$ where $V_{i, j}$ is a boolean value indicating whether the $j$-th branch has been executed at least once by the test case set of trial $i$ and $N_B$ is the number of branches in the target program.
The branch coverage value of trial $i$, $C_i$, is defined as the total number of branches covered (i.e., $C_i = \Sigma_{j=1}^{N_B} {V_{i, j}}$).

Among the 148 parameters available in KLEE, we included 61 parameters (i.e., $N_P$ = 61) in the parameter space by filtering out those that our domain experts deemed unimportant (e.g., ``-\--help'', ``-\--version''); see the supplementary material for details.
The number of trials in an experiment, $N_T$, typically ranges from a few hundred to a few thousand, with each trial limited to two minutes.
We used widely adopted benchmark programs such as gawk ($N_B=10,720$), gcal ($N_B=15,799$), and grep ($N_B=8,225$).

Given the large number of trials generated in an experiment, analyzing them individually is impractical.
To address this, we introduce the concept of a \textbf{trial group}, a subset of trials selected based on shared characteristics, such as similar branch coverage vectors or common parameter values.
This abstraction enables collective analysis at scale.
Formally, a trial group $g = \{i_1, ..., i_k\} \subset [1, N_T]$ is a subset of trials.
To quantify how exhaustive its test cases are, we define its \textbf{accumulated branch coverage value}, denoted as $accum(g)$.
First, we define its branch coverage vector as $V_g = \bigvee_{i \in g} V_i$ where $\bigvee$ denotes the element-wise logical OR operation. 
Then, $accum(g)$ is defined as the total number of unique branches covered (i.e., $accum(g) = \sum_{j=1}^{N_B} V_{g,j}$).
We define two trial groups as \textbf{complementary} if the accumulated branch coverage increases when the two groups are merged. 
Formally, given two trial groups ($g_1$, $g_2$), their complementarity $com(g_1, g_2)$ is defined as the coverage gain, $accum(g_1 \cup g_2) - max(accum(g_1), accum(g_2))$.

\edit{
Figure~\ref{fig:workflow} depicts the human-in-the-loop parameter tuning process supported by \system.
While conventional automated tuning operates within a fixed parameter space, \system enables users to analyze the outcomes of automated runs and iteratively refine the parameter space to improve efficiency and robustness, for example, by excluding ineffective parameters, removing failure-inducing values, and adjusting default settings.
}
\begin{figure*}[!t]
  \centering
  \includegraphics[width=0.81\linewidth ]{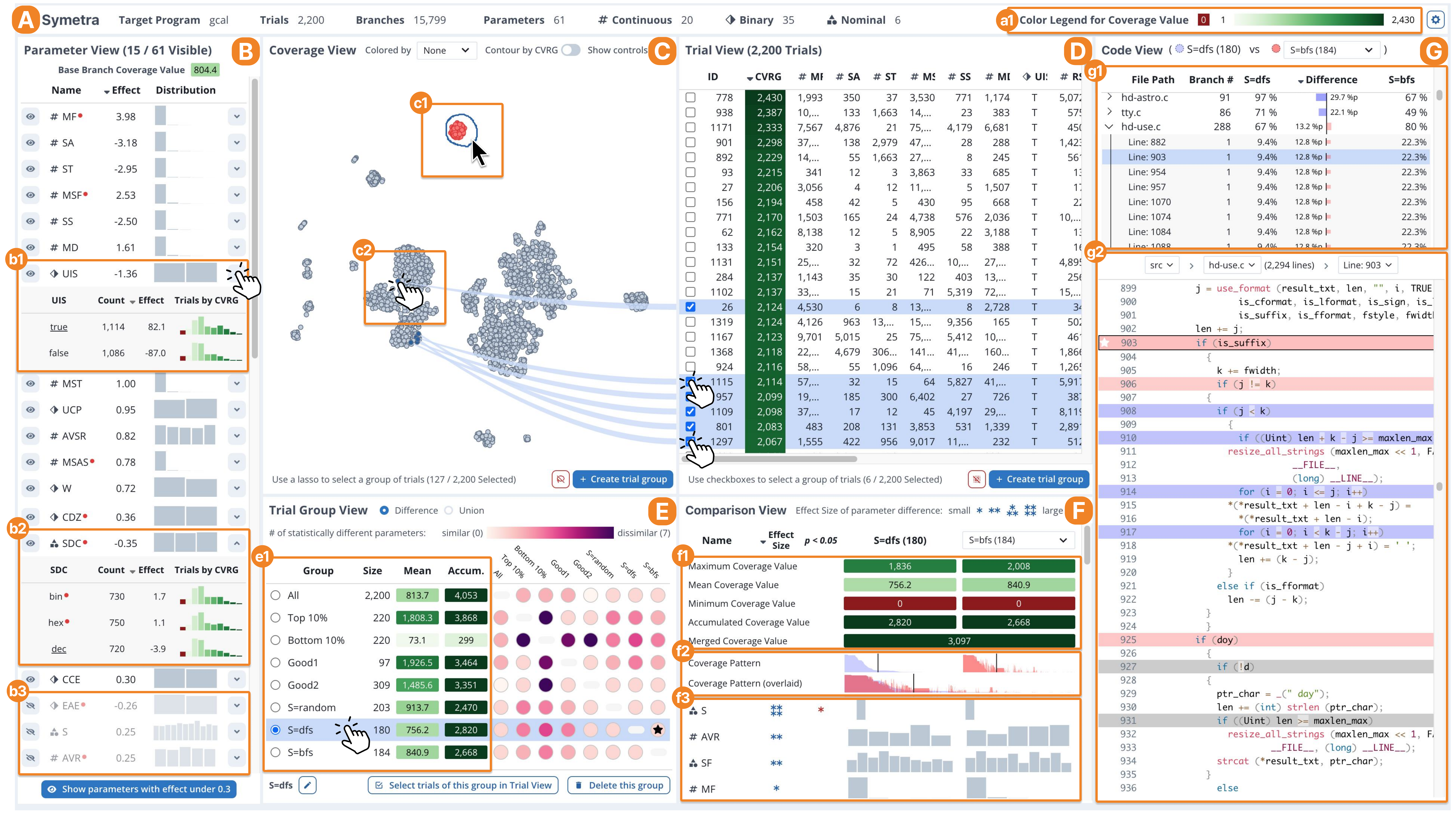}
  \caption{The interface of \system. The header (A) shows statistics and the legend for branch coverage values.
  Two overviews, the Parameter View (B) and the Coverage View (C),  provide overviews of parameters (i.e., the input to the engine) and branch coverage (i.e., the output from the engine).
  The user is making a trial group in the Trial View (D), and the Trial Group View (E) provides a summary of created trial groups, which could be compared through the Comparison View (F) and Code View (G). (\texttt{gcal}, $N_B=15,799$, $N_T=2,200$)
  }
  \label{fig:teaser}
  \vspace{-5mm}
\end{figure*}

\vspace{-2mm}

\subsection{User Goals and Tasks}



Our target users are researchers developing novel tuning algorithms to improve branch coverage and reduce the time required for test case generation.
They have two goals: to understand the parameter configurations produced by the tuner and their impact on branch coverage (\textbf{G1}), and to improve branch coverage by identifying complementary parameter values or configurations (\textbf{G2}).

During the collaboration, we observed critical inefficiencies in existing workflows for achieving these goals. In the absence of visual support, the user relied on ad hoc practices: manually parsing \textit{gcov} files (log files used to measure branch coverage) into spreadsheet programs, aggregating results in ways that obscured important details, and manually mapping code lines to their IDs. These practices fail to address the four \edit{critical limitations} outlined in Section 1. We identified five user tasks as follows:

\textbf{T1. Identifying the parameters and their defaults with the greatest effects on branch coverage.} 
The experts' primary task was to identify which parameters and their specific values had a significant positive or negative effect on branch coverage.
They were particularly interested in ranking parameters by significance to prioritize a smaller subset of parameters in the tuning process for efficiency.
They were also interested in comparing the performance of parameter values against the default values, recognizing that the default configuration might be suboptimal for the target program.

\textbf{T2. Exploring the branch coverage vectors of the tested configurations.}
Beyond identifying the effect of individual parameters, the experts aimed to analyze how the tested configurations covered branches in a program. Different configurations covering similar sets of branches could indicate redundancy. 
E1 highlighted this, stating \textit{``Understanding configurations that cover similar program branches can help us avoid wasting computational resources on redundant explorations.''}
Conversely, a configuration with a relatively low branch coverage value could still be deemed important if it exercises branches rarely covered by other configurations.

\textbf{T3. Clustering similar trials.}
Another task is to cluster similar trials, as they experienced that individually analyzing thousands of trials in an experiment was ineffective.
They wanted to cluster the trials into trial groups based on (1) branch coverage value, (2) branch coverage vectors, and (3) the value of a specific parameter.

\textbf{T4. Comparing trial groups.}
The experts aimed to understand how trial groups differed from each other in terms of parameter configurations and branch coverage vectors, which could explain the differences in branch coverage.
For instance, E2 remarked, \textit{``If we compare two high-performing groups, we want to understand why they are both effective but potentially in different ways.''}

\textbf{T5. Evaluating the impact of merged trial groups.}
The experts aimed to investigate whether merging the generated test case sets from different configurations could improve branch coverage. Specifically, they sought to evaluate the improvement in branch coverage achieved by merging test case sets from multiple trials.

\vspace{-1mm}

\section{The \system System}

\subsection{Design Goals}
During the design process of \system, we established four design goals to support the user tasks effectively as follows:

\textbf{DG1. Provide an overview of the parameters.}
Our first goal is to design a comprehensive overview that allows the experts to effectively identify and analyze the influence of parameters on branch coverage.
This overview should present the essential information needed to accomplish \textbf{T1}, including the magnitude of a parameter's impact, the effect of each parameter value on branch coverage, and a comparison of their performance relative to the default value.

\textbf{DG2. Facilitate exploration of branch coverage vectors.}
Our second design goal is to provide an overview of branch coverage vectors ($V_i$) to help identify similar or distinct coverage patterns (\textbf{T2}), which cannot be revealed by simple branch coverage values ($C_i$).
For instance, two trials might have the same branch coverage value while covering entirely different parts of the program.

\textbf{DG3. Support collective and comparative analysis of trials.}
To effectively handle a large number of trials, we aim to provide mechanisms for creating and managing trial groups (\textbf{T3}). 
We also support a comparative analysis between these trial groups (\textbf{T4}) and show the increase in the branch coverage value if we merge their test cases (\textbf{T5}).

\textbf{DG4. Enable tuner-agnostic analysis.}
While we used a certain tuner for this study, our experts also want to analyze the results of other tuners, such as Learch~\cite{10.1145/3460120.3484813} and ParaDySE~\cite{9507083}.
We opted for a tuner-agnostic approach to ensure broader applicability rather than focusing on a specific tuning algorithm.
As a result, we treat the tuner as a black box and visualize the relationships between parameter configurations and branch coverage, independent of the underlying tuning algorithm.

\vspace{-3mm}
\subsection{User Interface}

The user interface of \system consists of six views: Parameter View, Coverage View, Trial View, Trial Group View, Comparison View, and Code View (Figure~\ref{fig:teaser}).
The system provides two types of overviews; the Parameter View offers an overview of parameters and their effects (i.e., the input to the engine), while the Coverage and Trial Views provide an overview of branch coverage and trials (i.e., the output from the engine).
The remaining views support comparative analysis on trial groups. 
The header (Figure~\ref{fig:teaser}-A) displays simple statistics of the experiment, including the number of trials, branches, and parameters of each type.
We use a consistent colormap (Figure~\ref{fig:teaser}-a1) for branch coverage values across all views, with dark red indicating failed trials.

\vspace{-2mm}

\subsubsection{Parameter View}

The Parameter View provides an overview of the effect of the parameters on branch coverage values (Figure~\ref{fig:teaser}-B, \textbf{DG1}), supporting \textbf{T1}.
To gauge the importance of each parameter, we fit an XGBoost~\cite{10.1145/2939672.2939785} model as a surrogate model to approximate the relationship between parameter values and branch coverage.
We initially considered more transparent approaches, such as fitting a linear regression model and using its coefficients for parameter importance. Still, we found out that this approach was not capable of capturing the non-linear relationship between parameters, resulting in poor goodness of fit. 
In our empirical testing, XGBoost consistently outperformed other models in terms of performance and computational efficiency; see the supplementary material for details.

After fitting the model, we use SHAP (Shapley Additive exPlanations) values~\cite{10.5555/3295222.3295230} to quantify the contribution of each parameter to branch coverage.
These values represent the average contribution of each parameter value to the overall branch coverage, which can be added to the base coverage value (shown at the top of the Parameter View, Figure~\ref{fig:teaser}-B).
The effect of a parameter is computed as the weighted mean of the effects of its individual values, where the weight is the number of trials using each value.
For example, if the parameter MF has an effect of +3.98 in Figure~\ref{fig:teaser}-B, its values, on average, contributed an additional 3.98 branches to the base coverage value of 804.4.

\vspace{-2mm}

\paragraph*{Visual Encoding.} 
The Parameter View lists parameters in a tabular format, with each row displaying the type, name, and effect of a parameter, ordered by magnitude (\textbf{T1}).
The rightmost column shows a histogram of parameter value frequencies.
The user can click on a row to see the effect of each parameter value (Figure~\ref{fig:teaser}-b1). 
The default value is underlined, and values outperforming the default are marked with a red dot (Figure~\ref{fig:teaser}-b2).
The user can also control the visibility of parameters through checkboxes, which is linked to the Trial View (Figure~\ref{fig:teaser}-b3).
By default, parameters with an absolute effect below 0.3 are hidden.

\vspace{-1mm}

\begin{figure}[!t]
  \centering
  \includegraphics[width=0.8\linewidth]{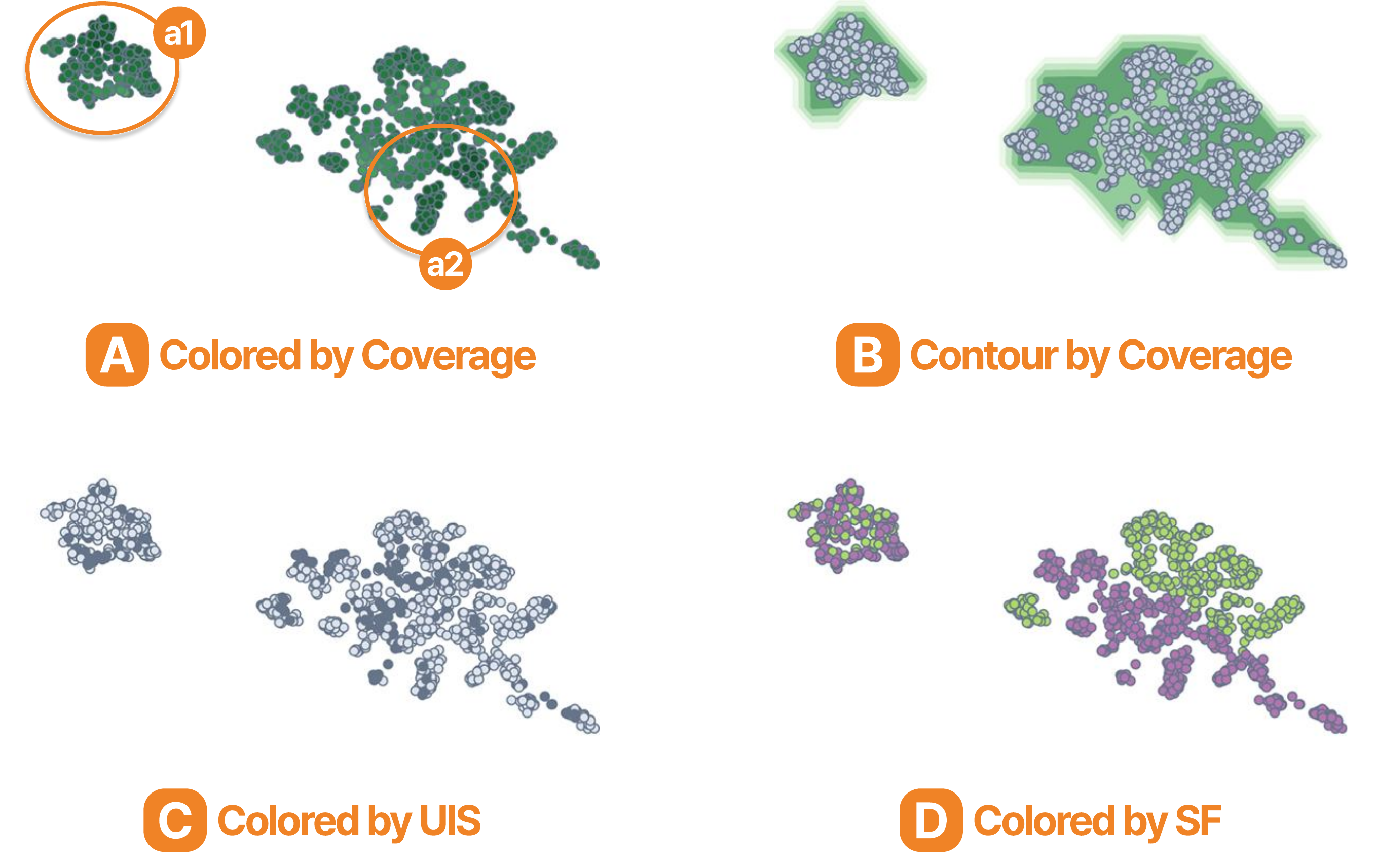}
  \vspace{-2mm}
  \caption{Branch coverage values are shown through point coloring (A) or density map contours (B) using the global colormap (Figure~\ref{fig:teaser}-a1). Points can be color-coded by a selected parameter to analyze coverage-parameter correlations, with color schemes adapting to parameter types: UIS (C) and SF (D). (\texttt{grep}, $N_B=8,225$, $N_T=2,200$)
 }
  \label{fig:coverage}
  \vspace{-5mm}
\end{figure}

\subsubsection{Coverage View}

\paragraph*{Visual Encoding.} The Coverage View provides an overview of branch coverage vectors of trials (Figure~\ref{fig:teaser}-C, \textbf{DG2}), supporting \textbf{T2}.
We project trials with similar branch coverage vectors into nearby positions using UMAP~\cite{McInnes2018} with Jaccard similarity (i.e., $dist(i,j) = 1 - |V_i \cap V_j|/|V_i \cup V_j|$), enabling the user to identify clusters of trials that cover similar branches (Figure~\ref{fig:coverage}-a1).
We use a conventional scatterplot with each point representing a trial whose coordinates are determined by UMAP.
The user can also use different dimensionality reduction techniques or distance measures, configurable in the interface. 

The user can color the points by coverage values (Figure~\ref{fig:coverage}-A) or use overlaid contours (Figure~\ref{fig:coverage}-B).
For example, in Figure~\ref{fig:coverage}, a1 and a2 both show high coverage values but appear separated, indicating they test different branches.
To support correlation analysis between parameter values and coverage vectors, we also allow the user to color-code points based on a selected parameter (Figure~\ref{fig:coverage}-C and D).
The user can select trials using the lasso tool (Figure~\ref{fig:teaser}-c1) and group them by clicking the ``Create Trial Group'' button (\textbf{T3}) for collective analysis (\textbf{T4}).
Clicking on a point scrolls the Trial View to the corresponding row, with explicit visual links for navigation (Figure~\ref{fig:teaser}-c2).

\paragraph*{Design Alternative.} 
While parallel coordinates plots (PCPs) are commonly used in HPO visualization systems~\cite{9222338, nni2021, 10.1145/3097983.3098043}, we opted against PCP for two reasons.
First, our experiments involve tens of parameters and thousands of trials, resulting in severe visual clutter.
Second, PCPs are designed to reveal similar configurations, whereas our experts focused on complementary configurations (\textbf{T5}) with dissimilar coverage vectors, which are not effectively shown when results are summarized on a single axis.


\subsubsection{Trial View}
\paragraph*{Visual Encoding.}
The Trial View provides the details of individual trials in a tabular format (Figure~\ref{fig:teaser}-D).
Each row represents a trial, showing its ID, color-coded branch coverage value ($C_i$), and configuration.
The user can select trials by checking checkboxes and group them using the ``Create Trial Group'' button (\textbf{T3}).
The view is linked to the Coverage View; selected rows are highlighted with connection marks in both views.


\definecolor{MineShaft}{rgb}{0.2,0.2,0.2}
\begin{table*}[!t]
\centering
\begin{tblr}{
  cell{1}{2} = {fg=MineShaft},
  cell{1}{3} = {fg=MineShaft},
  cell{1}{4} = {fg=MineShaft},
  cell{1}{5} = {fg=MineShaft},
  cell{3}{1} = {r=3}{},
  cell{6}{1} = {r=4}{},
  hline{1,10} = {-}{0.08em},
  hline{2-3,6} = {-}{},
  colspec = {Q[l,m,wd=1.3cm] Q[l,m,wd=0.8cm] Q[l,m,wd=9cm] Q[l,m,wd=1.7cm] Q[l,m, wd=3cm]},
  width = \linewidth,
}
\textbf{Type} & \textbf{Name}  & \textbf{Description} & \textbf{Default} & \textbf{Values / Ranges}\\
Binary & UIS  & Use constraint independence & true & \{true, false\}\\
Continuous & MF & Only fork this many times. Set to -1 to disable & 4096 & {[}0, 350,000]\\
 & ST & Amount of time to dedicate to seeds, before normal search & 5 & {[}1, 700,000]\\
 & SA & Replace by a symbolic argument with length N. & - & {[}1, 600]\\
Nominal & S & Specify the search heuristic & random-path & \{dfs, ..., random-path\}\\
 & SF & .ktest file to be used as seed & - & \{1, 2, 3, ..., 10\}\\
 & SDC  & Sets how bitvector constants are written in generated SMT-LIBv2 files & dec & \{dec, hex, bin\}\\
 & ST  & Select the implementation of switch & internal & \{internal, simple, llvm\}
\end{tblr}
\caption{
Examples of 8 out of the 61 parameters in KLEE with their names, full names, descriptions, default values, and possible value ranges, categorized by type: Binary (35 parameters) indicating a parameter with one of the boolean values, Continuous (20 parameters) for a parameter in a continuous range, and Nominal (6 parameters) indicating a categorical parameter whose value is one of the given set.
}
\label{tab:param}
\vspace{-5mm}
\end{table*}

\subsubsection{Trial Group View}
Serving as a starting point of collective and comparative analysis (\textbf{DG3}), the Trial Group View shows trial groups that the user has created on the Coverage and Trial Views (Figure~\ref{fig:teaser}-E).
To facilitate onboarding to analysis, we provide three default groups: \g{All} group consisting of all trials, and the \g{Top 10\%} and \g{Bottom 10\%} groups, containing 10\% of the trials with the highest and lowest branch coverage values, respectively.


\vspace{-2mm}

\paragraph*{Visual Encoding.}
We juxtapose a table and a heatmap where each row represents a trial group.
The table shows the name, size, mean, and accumulated branch coverage value (i.e., $accum(g)$) of each group (Figure~\ref{fig:teaser}-e1). 
The heatmap visualizes pairwise relationships between trial groups in two modes: difference mode encodes the number of statistically different parameters (\textbf{T4}), while union mode shows complementarity $com(g_1, g_2)$ (\textbf{T5}).

The user can click on a row to set it as the \textbf{active group} (Figure~\ref{fig:teaser}-e1) for the Comparison and Code Views, or click on a heatmap cell to compare the corresponding two groups.
Hovering over a row highlights its trials in the Coverage View.
We provide tooltips in the heatmap and the interactions for managing trial groups, e.g., naming or removing a group, and reselecting the trials of the active group within the Trial View.

\vspace{-2mm}

\paragraph*{Design Alternative.}
During the design process, we considered using a node-link diagram, where each node represents a trial group, with the thickness of a link indicating group similarity based on parameter values.
However, expert feedback revealed several limitations with this approach: it is not space-efficient, and the thickness of links is not perceptually effective. As a result, we opted for a heatmap visualization.

\subsubsection{Comparison View}
The Comparison View is designed to facilitate the comparison between trial groups, which can explain their difference in branch coverage (Figure~\ref{fig:teaser}-F, \textbf{DG3}). 
We support a pairwise comparison between two trial groups (\textbf{T4}): the active group (selected via the Trial Group View), and another chosen via a dropdown in the Comparison View.
The two groups are shown as juxtaposed columns for side-by-side comparison.

\vspace{-2mm}

\paragraph*{Visual Encoding.}
We support three types of comparative analysis (Figure~\ref{fig:teaser}-f1-f3).
First, branch coverage statistics (maximum, minimum, mean, accumulated) are shown for each group, along with the merged coverage value $accum(g_1 \cup g_2)$ to indicate if the two groups are complementary (\textbf{T5}).
For example, in Figure~\ref{fig:teaser}-f1, merging two groups increases coverage from 2,820 and 2,668 to 3,097, indicating their test cases could be merged to improve branch coverage complementarily.
Second, coverage pattern bar charts show the covering frequency of each branch, with overlaid charts highlighting differences between groups (Figure~\ref{fig:teaser}-f2).
Each bar represents a branch, ordered by covering frequency in the active group.
Clicking on a bar navigates to the corresponding line in the Code View.
Finally, parameter distributions are compared side-by-side as histograms, with effect size and statistical significance indicated by asterisks (Figure~\ref{fig:teaser}-f3).
We encode the magnitude of difference using blue asterisks and show a red asterisk if the difference is statistically significant.
See the supplementary material for details.

\subsubsection{Code View}

The Code View enables the user to compare the branch coverage of two trial groups at both the file and code levels (Figure~\ref{fig:teaser}-G, \textbf{DG3}). It is designed to help determine whether the two trial groups are complementary in their coverage (\textbf{T5}).
It consists of the file panel (Figure~\ref{fig:teaser}-g1) and the code panel (Figure~\ref{fig:teaser}-g2). 
\vspace{-2mm}

\paragraph*{Visual Encoding.} 
The file panel organizes source files in a hierarchical structure, allowing users to expand each file to inspect its internal branches and line-level details.
We define two coverage metrics: \textbf{file coverage}, the ratio of branches in a source file that are tested at least once by a given group, and \textbf{line coverage}, the average proportion of trials within a group that cover the branches on each line.
Both metrics are visualized as diverging bar charts placed next to each file or line; longer bars indicate stronger coverage bias toward one group, helping users identify files or lines that are predominantly tested by that group (Figure~\ref{fig:teaser}-g1).

In the code panel, differences in line coverage are visualized directly on the source code: each line's background color encodes the magnitude and direction of coverage difference between groups (Figure~\ref{fig:teaser}-g2).
For example, lines 903 and 906 (red) were covered more by \g{S=bfs}, while lines 908, 910, and 914 were covered more by \g{S=dfs}. Lines with no difference appear gray.
The panels are linked, and clicking on a file or line navigates between them. Branches selected in the Comparison View are highlighted with a black border.

\vspace{-2mm}

\subsection{Implementation}

\begin{figure*}[!t]
  \centering
  \includegraphics[width=0.7\linewidth]{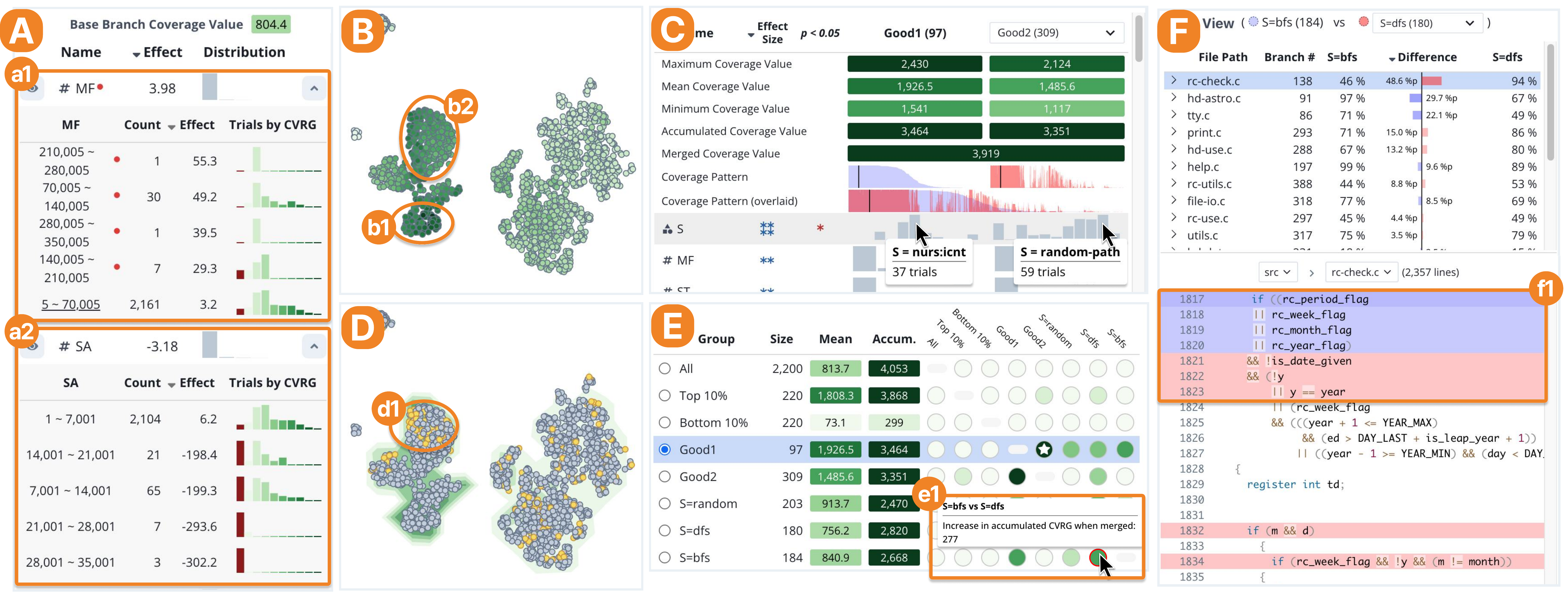}
  \caption{
Using the Parameter View (A), P1 examined the effects of the MF and SA parameters.
In the Coverage View (B), he identified two groups, \g{Good1} and \g{Good2}, that exhibited high overall coverage and similar coverage vectors.
The Comparison View (C) revealed that these groups had complementary values for S.
By hovering over \g{S=random}, he observed that most trials were located in low-coverage regions, though some overlapped with \g{Good2} (D).
Based on the trial groups created in the Trial Group View (E), he initiated a comparative analysis in the Code View, focusing on specific parameter values such as \val{dfs}, and \val{bfs} (F).
   (\texttt{gcal}, $N_B=15,799$, $N_T=2,200$)}
  \label{fig:case1}
  \vspace{-5mm}
\end{figure*}

\system was implemented with React.js~\cite{reactReact} using d3.js~\cite{10.1109/TVCG.2011.185} and visx~\cite{visx} to build visualization components. 
The source code and a web demo are available at \url{https://github.com/skku-idclab/Symetra}.
Our system requires five input files: 1) a JSON file specifying parameters, including names, types, default values, and descriptions (accessible via tooltips), 2) a CSV file with configurations and branch coverage values for each trial, 3) a text file having the set of branch IDs covered by each trial, 4) a JSON file indicating the line number and file name at which a covered branch is located, and 5) the source code of the target program.
This input structure has been verified as universally applicable across different tuners, enabling tuner-agnostic analysis (\textbf{DG4}).

\vspace{-2mm}
\section{Evaluation}
We evaluated \system using both qualitative and quantitative methods: two case studies examining how experts interact with the system, interviews with two extra experts to broaden the feedback, and a Human-In-the-Loop optimization study quantifying the system's impact on tuning performance in terms of efficiency and coverage improvements.

\vspace{-2.5mm}

\subsection{Case Study}
We conducted two case studies to evaluate whether and how \system can assist experts in achieving \textbf{G1} and \textbf{G2}.
Although we initially considered a comparative evaluation, the absence of dedicated tools in this context limited such an approach. We believe \system can therefore serve as a baseline for future research and instead conducted two case studies.
We invited two domain experts, P1 and P2, who each had eight years of experience in software engineering and testing, and both held Ph.D. degrees in software engineering. 
Neither expert was involved in the design process of our system.
P1, a researcher at a company's lab, specialized in automated software testing and focused on developing algorithms for automated bug fixing. 
P2, a researcher at an AI chip company, had experience generating test cases for a compiler.
We selected two widely used benchmark programs in software engineering: \texttt{gcal} ($N_B=15,799$) and \texttt{grep} ($N_B=8,225$). SymTuner~\cite{10.1145/3510003.3510185} was run for each program over three days, resulting in two datasets, each comprising 2,200 parameter configurations.

During the case studies, the experts explored one of the datasets on our system \edit{in a 90-minute session.}
The session began with a 20-minute interview to understand the participants’ experience in software testing and to introduce our system. 
For the next 60 minutes, the participants used the system to freely explore the dataset, analyzing parameter configurations and their impacts (\textbf{G1}) and seeking ways to improve branch coverage (\textbf{G2}). 
In the final 10 minutes, we collected subjective feedback through semi-structured interviews to gather insights into their experience using our system.
The study protocol was approved by the Institutional Review Board (IRB).

\vspace{-1mm}

\subsubsection{P1: Discovering Complementary Search Strategies}


\paragraph*{The Parameter View identified influential parameters and suboptimal defaults (T1).} 
P1 first observed that MF (maximum forks) exhibited the highest positive effect of +3.98, with values exceeding the default, improving coverage by up to +55.3 branches (Figure~\ref{fig:case1}-a1). 
He explained that coverage typically improved with higher fork limits when sufficient computational resources were available. In contrast, SA (symbolic arguments) showed negative effects when values exceeded 7,000 (Figure~\ref{fig:case1}-a2), which P1 attributed to solver overload and combinatorial explosion. The red dot indicators also revealed that SDC's default value was suboptimal (Figure~\ref{fig:teaser}-b2). Reflecting on this, he remarked that having visibility into these parameter effects would significantly assist in defining an effective parameter space.

\vspace{-3mm}

\paragraph*{The Coverage View revealed distinct clusters with different branch compositions (T2).} 
P1 identified two clusters and grouped them as \g{Good1} (97 trials) and \g{Good2} (309 trials) using the lasso tool (\textbf{T3}, Figure~\ref{fig:case1}-b1, b2). 
Although both achieved high coverage, their separation suggested they covered different branches. 
In the Trial Group View, P1 was surprised to find only one parameter differed significantly between them (\textbf{T4}).
Further investigation in the Comparison View identified S (search heuristic) as the differentiating factor. \g{Good1} used \val{nurs:icnt} while \g{Good2} used \val{random-path}. 
The Comparison View confirmed their complementarity: accumulated coverage of 3,464 and 3,351, respectively, increased to 3,919 when merged (\textbf{T5}, Figure~\ref{fig:case1}-C). 
Interestingly, P1 had expected trials with \val{random-path} to be uniformly distributed, as \val{random-path} explores execution paths stochastically, but found them cohesive in a specific region (Figure~\ref{fig:case1}-d1).
He hypothesized that discovering certain critical branches by chance substantially increased coverage, causing these trials to cluster.

\vspace{-3mm}

\paragraph*{The Code View verified strategy-specific patterns at the code level (T5).} 
To further investigate search strategy complementarity, P1 created groups for \g{S=dfs} and \g{S=bfs} (\textbf{T3}) and found their merged coverage increased from 2,820 and 2,668 individually to 3,097 combined (Figure~\ref{fig:case1}-e1). 
The Code View revealed the underlying reason: in \textit{rc-check.c}, \g{S=bfs} covered branches up to depth 4, while \g{S=dfs} penetrated to depth 7 (Figure~\ref{fig:case1}-f1). 
This confirmed that the two strategies complement each other by targeting different nesting levels—a finding consistent with P1's prior knowledge that could inform parameter space design for other programs.

\begin{figure*}[!t]
  \centering
  \includegraphics[width=0.7\linewidth]{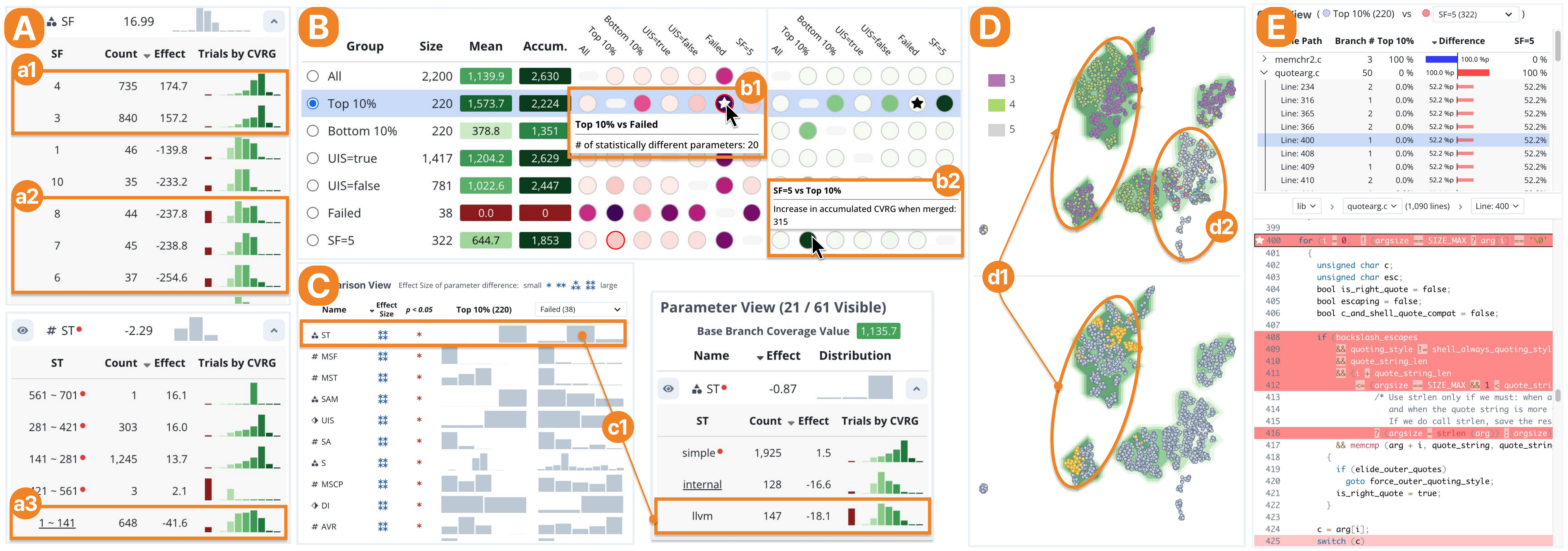}
   \caption{P2 examined the effects of SF and ST in the Parameter View (A).
   He chose a trial group of interest, \g{Top 10\%}, in the Trial Group View (B) and compared it with a group of failed trials, \g{Failed}, in the Comparison View (C), to investigate parameter values that possibly caused failures.
   In the Coverage View (D), he color-coded the trials (points) by SF and hovered over \g{Top 10\%} to highlight only the points in the group, exploring their relationship.
  The Code View (E) showed the trials in \g{SF=5} mainly covered \textit{quotearg.c}, which were related to handling inputs with quotation marks.
   (\texttt{grep}, $N_B=8,225$, $N_T=2,200$)}
  \label{fig:case2}
    \vspace{-5mm}
\end{figure*}

\vspace{-1mm}

\subsubsection{P2: Diagnosing Failures and Optimizing Seed File Selection }

\paragraph*{The Parameter View identified seed file values causing failures versus high coverage (T1).} 
P2 first focused on SF (seed file), which exhibited the highest effect of +16.99. Values 3 and 4 positively impacted coverage by up to +174.7 and +157.2 branches (Figure~\ref{fig:case2}-a1), which P2 attributed to special characters and regex patterns enabling comprehensive path exploration. Conversely, values 6, 7, and 8 resulted in frequent failures (Figure~\ref{fig:case2}-a2), likely due to invalid inputs causing premature termination. P2 also examined ST (seed time), finding that values near the default resulted in the lowest effect of -41.6 (Figure~\ref{fig:case2}-a3). He confirmed that values below 10 indicated insufficient seed execution time, while values above 140 would be more appropriate.

\vspace{-1mm}

\paragraph*{The Comparison View enabled systematic failure diagnosis (T4).} 
P2 noted that only 38 out of 2,200 trials failed, suggesting the tuner effectively avoided invalid inputs. To investigate what caused these failures, he created a \g{Failed} group in the Trial View (\textbf{T3}) and compared it with \g{Top 10\%} in the Trial Group View, finding that 20 parameters differed significantly (Figure~\ref{fig:case2}-b1). 
The Comparison View revealed that ST (switch type) set to \val{llvm} was a primary cause of failures. 
Of 38 failed trials, 29 used \val{llvm}, while \g{Top 10\%} frequently used \val{simple} (Figure~\ref{fig:case2}-c1).
This systematic comparison allowed P2 to pinpoint specific parameter values to exclude in future experiments.

\vspace{-1mm}

\paragraph*{The Coverage View revealed SF=5 as a potential complementary configuration (T2).} 
When P2 color-coded trials by SF, he discovered that \g{Top 10\%} mainly used values 3 and 4 (Figure~\ref{fig:case2}-d1), while SF=5 trials clustered separately despite not causing failures (Figure~\ref{fig:case2}-d2). He hypothesized these groups could be merged to improve coverage (\textbf{T5}). The Trial Group View confirmed that merging \g{SF=5} with \g{Top 10\%} increased accumulated coverage by 315 branches (Figure~\ref{fig:case2}-b2). 
The Code View revealed that \g{SF=5} uniquely covered \textit{quotearg.c}, which contained branches requiring quotation mark handling in regular expressions (Figure~\ref{fig:case2}-E). 
P2 explained that such patterns require specific input structures unlikely to emerge through mutation, concluding that an effective testing strategy might involve having different seed files for certain codes, as specific input structures are essential for comprehensive testing.

\vspace{-2mm}

\subsubsection{Results}
Across the two case studies, both experts effectively used \system's coordinated views to address distinct analytical goals.
P1 leveraged the Coverage View to identify complementary search strategies that were not apparent from aggregate branch coverage values alone, whereas P2 used the Parameter View to diagnose failure-causing parameter values.
Both case studies demonstrated that the drill-down workflow, which starts from global overviews (Parameter and Coverage Views) and proceeds to detailed comparisons (Comparison and Code Views), supported not only pattern discovery but also verification of their underlying causes at the code level.
In particular, the Comparison and Code Views enabled experts to examine differences between trial groups and confirm coverage disparities directly within source files.

Post-study interviews further indicated that both participants regarded the Comparison and Code Views as particularly effective for comparative analysis to improve branch coverage (\textbf{G2}), replacing manual and time-consuming workflows.
P1 highlighted the streamlined debugging workflow enabled by the Code View, stating:
\textit{``Previously, analyzing covered branches required at least three steps: identifying the branch ID, locating the file and line number, and opening the source in a code editor. \system reduces this to a single integrated process.''}

Both experts also emphasized the usefulness of the Parameter View for understanding how individual parameters and their values influenced branch coverage (\textbf{G1}).
The case studies revealed that effective configurations varied substantially across programs.
For \texttt{gcal}, MF (maximum forks) was the dominant factor influencing coverage, whereas for \texttt{grep}, SF (seed file) had the strongest effect.
This variation highlights the importance of program-specific analysis when tuning symbolic execution engines.

\vspace{-4mm}

\subsection{Expert Interviews}
To broaden feedback beyond the case studies, we interviewed two additional domain experts. P3 was a researcher interested in analyzing differences between human-generated and machine-generated test cases, while P4 was a researcher focused on developing optimization techniques for symbolic execution engines, particularly search strategies and constraint-solving algorithms. 
Neither expert was involved in the design process of our system.
Each interview lasted one hour: preliminary discussion (20 minutes), a system demonstration (20 minutes), and feedback (20 minutes).

Both experts agreed that the current practices for analyzing engine runs are inefficient; for instance, P3 reported that parsing and interpreting the results of his experiments can take several hours, the main bottleneck in his research.
P3 greatly appreciated the Coverage View for highlighting trials with similar coverage values but targeting different code regions \textbf{(G1)}. He remarked:
\textit{``The Coverage View instantly reveals patterns I spent hours searching for manually. Seeing trials cluster by branch coverage vectors can directly support my research on making symbolic execution engines generate more human-like test cases.''}
He further noted that this overview served as a starting point to drill down into the Code View, where he could examine differences in the tested code regions in more detail.
P4, on the other hand, discovered through the Parameter View that common KLEE configurations used in published research were suboptimal; by adjusting the defaults, he was able to increase coverage by more than 100 branches \textbf{(G2)}. P4 commented:
\textit{``Seeing the quantified parameter effects changes how I set up experiments. The default configurations I have been using were not actually optimal for my target programs, requiring careful reconsideration.''}
Their comments highlight the potential of \system to support more efficient workflows in ongoing research.

\vspace{-3mm}

\subsection{Human-in-the-Loop Optimization in Practice}

We conducted an experiment to quantify the extent to which Human-in-the-Loop tuning can further improve the branch coverage achieved by a fully automated approach.
E1 and E2 analyzed an initial tuner run and refined parameters to improve branch coverage.
After each adjustment, we re-ran the experiment to assess improvement over three iterations.
We used \texttt{gawk} ($N_B=10{,}720$) with SymTuner~\cite{10.1145/3510003.3510185}, generating 800 trials per experiment (27 hours each).
We measured three metrics: $Acc_{800}$ (the accumulated coverage of all 800 trials), $N_{failed}$ (the number of failed trials with zero coverage), and $Acc_{30}$ (the accumulated coverage of the first 30 trials, capturing early convergence).
The initial experiment\edit{, serving as the fully automated baseline,} achieved $Acc_{800}=3{,}832$, $N_{failed}=128$, and $Acc_{30}=3{,}411$.

\paragraph*{Filtering Out Parameters with Little Effects.}
In the first iteration, while examining the Parameter View, the experts chose to exclude 28 parameters with an average effect below 0.3.
They also noticed the parameters with values outperforming the default, highlighted by red dots in the Parameter View, and adjusted their defaults to the values with the greatest positive effect.
After the refinement, they achieved $Acc_{800}=4{,}093$ (6.8\% $\uparrow$), $N_{failed}=22$ (82.8\% $\downarrow$), and $Acc_{30}=3{,}651$ (7.0\% $\uparrow$).

\paragraph*{Removing the Parameter Values Commonly Used in the Failed Trials.}
In the second iteration, they created a trial group of the 22 failed runs (i.e., $C_i=0$) through the Trial View.
Inspecting the histograms of parameter values in the Comparison View, they identified four parameter values used in the group, which they believed to lead to failures: KCO=\val{true}, ST=\val{llvm}, DI=\val{true}, and DV=\val{false}. 
After excluding those values, they achieved $Acc_{800}=4{,}379$ (7.0\% $\uparrow$), $N_{failed}=0$ (100\% $\downarrow$), and $Acc_{30}=3{,}704$ (1.5\% $\uparrow$).

\paragraph*{Redefining the Defaults.}
In the third iteration, they found that there were 16 parameters with a better default (i.e., red dots) in the Parameter View, while they changed the defaults in the first iteration.
Considering the interaction effect of the changes made in the second iteration, they adjusted the defaults of the 16 parameters to the values that demonstrated the greatest positive effect.
After the third refinement, they achieved $Acc_{800}=4{,}502$ (2.8\% $\uparrow$), $N_{failed}=0$ (no change), and $Acc_{30}=3{,}813$ (2.9\% $\uparrow$). 
Notably, $Acc_{30}=3{,}813$ is comparable to $Acc_{800}=3{,}832$ from the initial experiment that required 27 hours.

\edit{Through three iterations, each guided by analytical evidence from \system, Human-in-the-Loop tuning achieved coverage of 3,813 within one hour, comparable to 3,832 achieved by the fully automated approach in 27 hours.}
\vspace{-2mm}



\section{Discussion and Limitations}

Our evaluations demonstrate that \system enables experts to understand parameter impacts (\textbf{G1}) and identify complementary configurations (\textbf{G2}).
The optimization experiment shows that insights from \system can transfer to future tuning; the refined parameter space achieved the initial 27-hour coverage within one hour, leaving the remaining time to explore additional branches. This improvement suggests that \system can accelerate testing workflows for similar programs, such as newer versions of the same software.

We learned two lessons from applying visual analytics to a domain where it has been rarely used, software testing.
First, engaging with domain experts revealed that users often overtrust default settings that have remained unexamined for over a decade, even though such defaults may not be tailored to their target programs. Visual analytics helped surface these hidden assumptions and motivated experts to adopt more effective parameter choices.
As parameter optimization is used across many computational domains, this line of research warrants broader exploration to support more informed and transparent tuning practices in other fields as well.

Second, during the design process, we found that our experts were reluctant to use complex visual encodings, such as parallel coordinates, which have been used in previous interactive HPO systems.
Because these experts are already cognitively loaded with demanding tasks, introducing unfamiliar or visually intricate representations can easily become overwhelming rather than helpful.
As a result, we prioritized simpler and more familiar representations that minimize interpretation overhead, such as bar charts, histograms, scatterplots, and tables.
However, as relying solely on simple visualizations may limit efficiency in the long run, it is important to explore ways to help experts gradually adopt more expressive visual encodings without overwhelming them.

We identify five limitations and directions for future work. 
First, our system visualizes the results of an experiment without further interaction with the tuner or engine. 
Integrating them would enable the user to design configurations, launch them, and immediately visualize new trials.
Second, we employed XGBoost and SHAP values to quantify parameter impacts. 
While effective, alternative methods could be explored, such as computing variance to highlight differences across parameter values.
\edit{Third, displaying six coordinated views simultaneously may increase the initial learning cost. Collapsible panels that progressively reveal views on demand could help new users focus on relevant subsets of the interface.
Fourth, our system begins analysis from the parameter space and drills down to code-level details. 
Supporting the other direction, such as starting from a specific code region to identify relevant parameters, or analyzing how parameter effects vary with experiment duration, remains an open challenge.
Lastly,} we engaged with six domain experts and three software programs, \edit{and the quantitative experiment used a single program,} which poses a potential threat to ecological validity.
Extending this Human-in-the-Loop approach to other automated tuning systems beyond symbolic execution could further validate its generalizability.

\vspace{-2mm}

\section{Conclusion}
We present \system, a visual analytics system designed to support the parameter tuning process of symbolic execution engines. 
To address the challenge of managing a large number of configurations, our system supports collective analysis, enabling the user to compare trial groups and examine differences in coverage at both the file and code levels.
Our case studies and expert interviews reveal that experts were able to visually analyze the parameter tuning process, gaining insights into parameter impacts and coverage patterns. 
Finally, our optimization experiment demonstrates that insights from \system can iteratively improve branch coverage, showcasing the benefits of Human-in-the-Loop parameter tuning.

\vspace{-3mm}
\section*{Acknowledgements}



This work was supported by Institute of Information \& communications Technology Planning \& Evaluation (IITP) grant funded by the Korea government (MSIT) (RS-2019-II190421, Artificial Intelligence Graduate School Program (Sungkyunkwan University); RS-2024-00438686, Development of software reliability improvement technology through identification of abnormal open sources and automatic application of DevSecOps) and also by the National Research Foundation of Korea (NRF) grant funded by the Korea government (RS-2025-24873100).

\bibliographystyle{eg-alpha-doi} 
\bibliography{main}       




\end{document}